\documentclass[journal]{IEEEtran}
\usepackage{amsmath}
\usepackage{amssymb}
\usepackage{graphicx}
\usepackage{hyperref}
\usepackage{kotex}
\usepackage[table,xcdraw]{xcolor}
\usepackage{xpatch}
\usepackage{orcidlink}
\usepackage{fancyhdr}
\ifCLASSOPTIONcompsoc
    \usepackage[caption=false, font=normalsize, labelfont=sf, textfont=sf]{subfig}
\else
\usepackage[caption=false, font=footnotesize]{subfig}
\fi

\usepackage{cite}

\ifCLASSINFOpdf
\else
\fi

\usepackage{url}
\usepackage{hyperref}

\begin{document}
\title{Conditional Brownian Bridge Diffusion Model for VHR SAR to Optical Image Translation}
\author{Seon-Hoon~Kim\orcidlink{0009-0008-9261-6828},~\IEEEmembership{Graduate Student Member,~IEEE,}
        Daewon~Chung
\thanks{This research was supported by Korea Institute of Marine Science \& Technology Promotion (KIMST) funded by the Korea Coast Guard (RS-2023-00238652, Integrated Satellite-based Applications Development for Korea Coast Guard).}
\thanks{Seon-Hoon~Kim is with University of Science and Technology (UST), Daejeon, 34113, Republic of Korea (e-mail: egshkim@gmail.com).}
\thanks{Daewon~Chung is with the Korea Aerospace Research Institute, Daejeon 34133, Republic of Korea (e-mail: dwchung@kari.re.kr).}
\thanks{© 2025 IEEE. Personal use of this material is permitted. Permission from IEEE must be obtained for all other uses. DOI: \href{https://doi.org/10.1109/LGRS.2025.3562401}{10.1109/LGRS.2025.3562401}}}

\markboth{Journal of \LaTeX\ Class Files,~Vol.~13, No.~9, September~2014}%
{Shell \MakeLowercase{\textit{et al.}}: Bare Demo of IEEEtran.cls for Journals}


\maketitle

\begin{abstract}
Synthetic Aperture Radar (SAR) imaging technology provides the unique advantage of being able to collect data regardless of weather conditions and time. However, SAR images exhibit complex backscatter patterns and speckle noise, which necessitate expertise for interpretation. Research on translating SAR images into optical-like representations has been conducted to aid the interpretation of SAR data. Nevertheless, existing studies have predominantly utilized low-resolution satellite imagery datasets and have largely been based on Generative Adversarial Network (GAN) which are known for their training instability and low fidelity. To overcome these limitations of low-resolution data usage and GAN-based approaches, this letter introduces a conditional image-to-image translation approach based on Brownian Bridge Diffusion Model (BBDM). We conducted comprehensive experiments on the MSAW dataset, a paired SAR and optical images collection of 0.5m Very-High-Resolution (VHR). The experimental results indicate that our method surpasses both the Conditional Diffusion Models (CDMs) and the GAN-based models in diverse perceptual quality metrics.
\end{abstract}


\IEEEpeerreviewmaketitle

\section{Introduction}
\IEEEPARstart{N}{otwithstanding} the all-weather, day-and-night capability of Synthetic Aperture Radar (SAR), its advantages are accompanied by notable challenges. Inherent and unintentional artifacts such as speckle noise and geometric distortions complicate their analysis. To improve the interpretability of SAR imagery while preserving its all-weather and Very-High-Resolution capturing capabilities, VHR SAR to optical image translation is needed. Recent studies have explored deep learning methods to translate SAR imagery into optical-like images. This translation process becomes significantly more complex at higher resolutions due to increased detail complexity, greater computational demands, and a wider domain gap between SAR and optical imagery characteristics.

\begin{figure*}
    \centering
    \includegraphics[width=0.9\textwidth]{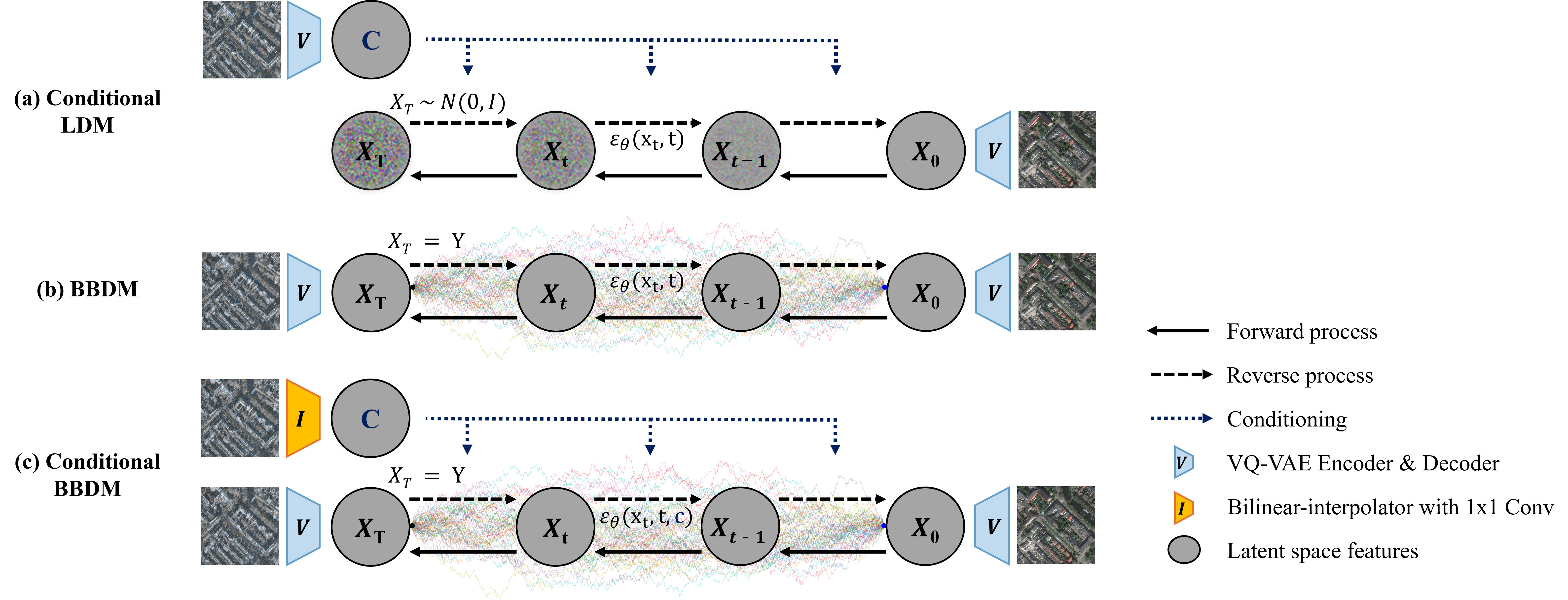}
    \caption{The directed graphical models of diffusion based methods. (a) Conditional LDM. (b) BBDM. (c) Conditional BBDM. $X_0$ denotes the latent features of optical imagery and Y denotes the latent features of SAR imagery. BBDM framework directly translates $X_0$ from $X_T$ through Brownian bridge. Otherwise, conditional LDM framework gradually reconstructs $X_0$ from noisy $X_T$, guided by the condition from SAR imagery. Conditional BBDM employs condition to guide the direct mapping from $X_T$ to $X_0$. Both (b) and (c) depict a Brownian bridge process in the background of the diffusion process.}
    \label{fig:Figure 1}
\end{figure*}

However, research on SAR to optical translation using Very High Resolution (VHR) data with sub-meter resolution is extremely scarce. Most existing studies have utilized datasets that fall short of sub-meter VHR standards \cite{SAR2OPT2018_GAN_ku2018method, SAR2OPT2022_improved_CGAN_SEN12, SAR2OPT2022_sar2color_GAN_SEN12, conditionalDiffusion_SAR2Opt_SEN12_GF3, conditionalDiffusion_SAR2Opt_with_Colorsupervised_SEN12}. The widely used SEN12 dataset \cite{SEN12} consists of paired SAR and optical images at 10-meter resolution. Although this dataset has been immensely valuable for various remote sensing applications, its relatively coarse resolution limits its applicability for VHR SAR to optical translation tasks. Other datasets such as WHU-OPT-SAR \cite{WHU-OPT-SAR-dataset} and SARptical \cite{wang2018sarptical}, while offering improvements in resolution, still do not meet the sub-meter resolution criteria for true VHR data.

Yet, even with this inherent advantage of utilizing these low-resolution images, existing GAN-based approaches for translating SAR-to-optical have struggled to achieve practical performance, facing issues like training instability, mode collapse, and geometric loss with complex scenes \cite{SAR2OPT2018_GAN_ku2018method, SAR2OPT2022_improved_CGAN_SEN12, SAR2OPT2022_sar2color_GAN_SEN12}, \cite{SAR2OPT2022_GAN_MSAW_misused}. Only a few recent studies \cite{conditionalDiffusion_SAR2Opt_SEN12_GF3, conditionalDiffusion_SAR2Opt_with_Colorsupervised_SEN12} have explored Conditional Diffusion Models (CDMs) to overcome these limitations of GAN-based models. More recently, Shi et al. \cite{shi2024brain} proposed a brain-inspired approach that enhances CDMs by incorporating self-attention mechanisms and long-skip connections to improve feature extraction capabilities. CDMs currently dominate the image synthesis field \cite{diffusion_beat_GANs} including image-to-image translation. Still, CDMs suffer from difficulties with limited model generalization despite their potential. They lack robust theoretical foundations to ensure that the outcome accurately represents the intended conditional distribution \cite{BBDM}. These models often experience performance degradation when translating between significantly disparate domains.

We used the MSAW dataset \cite{MSAW_SpaceNet6} which provides overlapped pairs of 0.5m VHR SAR and optical imagery to alleviate the limitations in existing research. In contrast to the earlier study \cite{SAR2OPT2022_GAN_MSAW_misused} that used random train-validation splits on overlapping areas, we split the dataset based on longitude to ensure truly unseen validation data. We adopted an innovative image translation framework which is based on the self-attention incorporated UNet model, the Brownian Bridge Diffusion Model (BBDM) \cite{BBDM}, training on the carefully partitioned MSAW dataset. BBDM established a mathematical foundation for diffusion-based image translation. BBDM utilizes a bidirectional diffusion process based on stochastic Brownian bridge to directly learn the mapping between domains.

Basically, the original BBDM performs the diffusion process in a compressed latent space to operate it more efficiently by leveraging compact yet informative latent representations. While this approach offers advantages over GANs and conventional diffusion models through stable training with its Brownian Bridge mapping incorporated gradual denoising process, it lacks mechanisms for domain-specific guidance. Inspired by CDMs \cite{diffusion_beat_GANs}, we adjusted the BBDM framework for VHR SAR to optical image translation. Our approach incorporates bilinearly interpolated information from the pixel space as a condition, which fundamentally differs from the original BBDM's unconditioned approach. This ensures that the model receives spatial information to guide the translation process. This combination of latent representations and conditioned information not only maintains the stable training advantages of the original BBDM over GANs, but also overcomes the original BBDM's limitations by striking a balance between computational efficiency and the preservation of essential spatial details from the SAR image.

Experimental results show our conditional BBDM framework significantly improved SAR-to-optical translation quality. The proposed approach outperforms both GAN-based models and conditional Latent Diffusion Model (LDM) \cite{LDM} across various metrics. These results highlight the benefits of using conditions. This proves that our proposed method provides a robust framework for bridging the gap between SAR and optical imagery.

The main contributions in this letter include the following:

1)	We introduce a novel image-to-image translation framework (BBDM) to the SAR to optical research field. This offers an alternative to the predominantly used GAN models.

2)	In contrast to the original BBDM's unconditioned approach, by incorporating spatially interpolated information from the pixel space as a condition into BBDM, we achieved substantial improvements in both structural fidelity and visual quality for VHR SAR to optical imagery translation.

3)	We conducted SAR to optical image translation experiments on a 0.5m resolution VHR imagery dataset (MSAW) and demonstrated that the proposed model significantly outperforms both conditional LDM and existing GAN-based models across multiple metrics.

\section{METHODOLOGY}
\subsection{Conditional Latent Diffusion Model}
Recent progress in generative modeling has been driven by diffusion-based methods. Ho et al. \cite{DDPM} effectively implemented the diffusion framework for image synthesis. Rombach et al. \cite{LDM} proposed conditional LDM which performs the diffusion process in the compressed latent space as illustrated in Fig. \ref{fig:Figure 1}. Forward process gradually transforms an original latent feature $x_0$ into Gaussian noise $x_T \sim {N}(0, 1)$ through T iterative noise addition steps where T is typically equal to 1,000:

\begin{equation}
q(x_t|x_{t-1}) := \mathcal{N}(x_t; \sqrt{1 - \beta_t}x_{t-1}, \beta_t \textbf{\textit{I}})
\end{equation}

where $\beta_t$ is a linearly scheduled variance. As shown in Fig. \ref{fig:Figure 1}, the reverse process reconstructs the original image by estimating noise at each step:

\begin{equation}
p_\theta(x_{t-1}|x_t) := \mathcal{N}(x_{t-1}; \mu_\theta(x_t, t), \sigma_t^2\textbf{\textit{I}})
\end{equation}

where $\sigma_t^2I$ represents the time-dependent constants which are not trainable. For image-to-image translation, conditional LDM incorporate source domain images as conditions, typically through cross-attention or concatenation. We can modify the reverse process notation as:

\begin{equation}
p_\theta(x_{t-1}|x_t, c) := \mathcal{N}(x_{t-1}; \mu_\theta(x_t, t, c), \sigma_t^2\textbf{\textit{I}})
\end{equation}

Conditional LDM are trained to optimize a variational lower bound, learning to reverse the gradual noise accumulation process for high-quality data generation \cite{LDM}. For inference, conditional LDM starts from a complete Gaussian noise and progressively denoises it guided by the given condition.

\subsection{Brownian Bridge Diffusion Model}

Li et al. \cite{BBDM} proposed Brownian Bridge Diffusion Model (BBDM) which is a novel approach for image translation grounded in concepts from Brownian bridge process. A Brownian bridge represents a continuous-time stochastic process with fixed start and end points. BBDM applies this concept to represent the translation as a probabilistic transformation between two fixed states, as shown in Fig. \ref{fig:Figure 1}. For a given starting point $x_0$ and ending point $x_T$, the probability distribution of intermediate states $x_t$ in a Brownian bridge process is formulated as:

\begin{equation}
p(x_t|x_0, x_T) = \mathcal{N}(x_t; (1 - \frac{t}{T})x_0 + \frac{t}{T}x_T, \frac{t(T-t)}{T} \textbf{\textit{I}})
\end{equation}

To downscale the maximum variance that peaks at the midpoint, $\delta_{\frac{T}{2}} = \frac{T}{4}$, \cite{BBDM} formulate the Brownian bridge forward diffusion process as:
\begin{equation}
\label{Forward diffusion process of Brownian bridge}
q_{BB}(x_t|x_0, y) = \mathcal{N}(x_t; (1 - m_t)x_0 + m_t y, \delta_t \textbf{\textit{I}})
\end{equation}

where $x_0$ is the latent representation of the initial image and $y$ denotes the latent representation of the target domain image, $m_t = t/T$, and $\delta_t = 2(m_t - m_t^2)$ which is the scheduled variance. The  maximum variance $\delta_{max} = \delta_{\frac{T}{2}}$ becomes $\frac{1}{2}$ in this formulation, . When $t=0$, $m_0$ is equal to 0, the mean and variance becomes $x_0$ and 0, respectively. When $t=T$, $m_T$ is equal to 1, the mean and variance becomes $y$ and 0, respectively. This formulation is analogous to the Brownian bridge, with $x_0$ and $y$ serving as the fixed endpoints.

By combining two equations of $x_t$ and $x_{t-1}$ defined from Eq.(\ref{Forward diffusion process of Brownian bridge}), the transition probability can be derived as proposed in \cite{BBDM}:
\begin{equation}
    \begin{aligned}
        q_{BB}(x_t|x_{t-1}, y) = \mathcal{N}(x_t; \frac{1-m_t}{1-m_{t-1}}x_{t-1} + \\ 
        (m_t - \frac{1-m_t}{1-m_{t-1}}m_{t-1})y, \delta_{t|t-1}\textbf{\textit{I}})
    \end{aligned}
\label{Transition probability of BBDM}
\end{equation}

where $\delta_{t|t-1}$ is derived as:
\begin{equation}
\delta_{t|t-1} = \delta_t - \delta_{t-1}\frac{(1-m_t)^2}{(1-m_{t-1})^2}
\end{equation}

The transition probability given by Eq.(\ref{Transition probability of BBDM}) still describes a direct mapping between two fixed points.

The reverse process aims to estimate the mean of the noise at each step:
\begin{equation}
p_\theta(x_{t-1}|x_t, y) = \mathcal{N}(x_{t-1}; \mu_\theta(x_t, t), \tilde{\delta}_t \textbf{\textit{I}})
\end{equation}

where $\tilde{\delta}_t$ represents the variance of the noise at each step:
\begin{equation}
\tilde{\delta}_t = \frac{\delta_{t|t-1} \cdot \delta_{t-1}}{\delta_t}
\end{equation}

We can express the training objective of BBDM, derived from the Evidence Lower Bound (ELBO), in a simplified form as proposed in \cite{BBDM}:

\begin{equation}
\mathbb{E}{x_0,y,\epsilon}[c_{\epsilon t}||m_t(y - x_0) + \sqrt{\delta_t}\epsilon - \epsilon_\theta(x_t, t)||^2]
\end{equation}

where $c_{\epsilon t}$ is defined as:

\begin{equation}
c_{\epsilon t} = (1 - m_{t-1}) \frac{\delta_{t|t-1}}{\delta_t}
\end{equation}

The formulation enables BBDM to learn a smooth mapping between image domains and result in higher quality translations. The application of the Brownian bridge concept ensures that the translation process is anchored at both the source and target images. It provides a natural framework for image-to-image translation.

Although BBDM has some advantages such as direct modeling of domain transitions and efficient handling of high-resolution images through latent space operations, it presents certain limitation. The primary constraint is the absence of explicit conditioning mechanisms. BBDM does not incorporate additional guiding information during the translation process. This limitation potentially restrict its utility in applications requiring precise control over the translation process.

\subsection{Conditional BBDM for SAR to optical}

We propose a conditional BBDM (cBBDM) to address the limitations of the original BBDM and adapt it for SAR to optical image translation. This modification allows for the incorporation of additional guiding information during the translation process. As a result, it produces more refined results.

As with the original BBDM, cBBDM mainly operates in the compressed latent space of images. It reduces computational complexity while preserving essential features. As illustrated in Fig. \ref{fig:Figure 1}, we introduce a conditioning variable $c$, derived from the pixel space of given SAR images to guide the translation process. This condition is obtained by applying bilinear interpolation to the SAR images to align with the 16-fold compressed latent space dimensions, followed by a 1x1 convolution. The condition $c$ is incorporated into the diffusion process through channel-wise concatenation ([,]):

\begin{equation}
p_\theta(x_{t-1}|x_t, x_T, c) = \mathcal{N}(x_{t-1}; \mu_\theta([x_t, c], t), \tilde{\delta}_t \textbf{\textit{I}})
\end{equation}

The training objective of cBBDM is modified to include this conditioning information:

\begin{equation}
\mathbb{E}{x_0,x_T,c,\epsilon}[c_{\epsilon t}||m_t(x_T - x_0) + \sqrt{\delta_t}\epsilon - \epsilon_\theta([x_t, c], t)||^2]
\end{equation}

where $\epsilon_\theta([x_t, c], t)$ is the noise prediction network that takes the concatenated representation of latent features and condition as input.

\begin{table*}[hbt!]
\caption{Performance Comparison of Different Models for VHR SAR to Optical Image Translation. The Directional Arrows adjacent to each Evaluation Metric Indicate the Metric's Correlation with Better Quality. The Highest score is Marked in Red and the Second-Highest score is Marked in Green}
\renewcommand{\arraystretch}{1.35}
\label{table:Quantitative comparison of different methods}
\centering
\resizebox{\textwidth}{!}{%
\begin{tabular}{c|c|c|c|c|c|c|c|c}\hline
\textbf{Methods} & \textbf{LPIPS-ALEX↓} & \textbf{LPIPS-VGG↓} & \textbf{LPIPS-SQUEEZE↓} & \textbf{FID↓} & \textbf{SAM↓} & \textbf{CHD↓} & \textbf{CW-SSIM↑} & \textbf{FSIMc↑} \\ \hline Pix2pix & 0.5298 & 0.6195 & 0.3690 & 241.58 & 0.0734 & 0.2865 & {\color[HTML]{329A9D} 0.4100} & {\color[HTML]{329A9D} 0.6532} \\ CycleGAN & 0.5395 & 0.6429 & 0.4254 & 222.95 & 0.0903 & 0.2961 & 0.4058 & 0.6396 \\ Conditional LDM & 0.6268 & 0.6333 & 0.4477 & 256.17 & 0.0882 & 0.5317 & 0.3568 & 0.6250 \\ BBDM & {\color[HTML]{329A9D} 0.4946} & {\color[HTML]{329A9D} 0.5952} & {\color[HTML]{329A9D} 0.3599} & {\color[HTML]{329A9D} 197.54} & {\color[HTML]{329A9D} 0.0728} & {\color[HTML]{329A9D} 0.2785} & 0.3905 & 0.6462 \\ Conditional BBDM (Ours) & {\color[HTML]{FD6864} \textbf{0.4579}} & {\color[HTML]{FD6864} \textbf{0.5744}} & {\color[HTML]{FD6864} \textbf{0.3286}} & {\color[HTML]{FD6864} \textbf{177.79}} & {\color[HTML]{FD6864} \textbf{0.0709}} & {\color[HTML]{FD6864} \textbf{0.2411}} & {\color[HTML]{FD6864} \textbf{0.4138}} & {\color[HTML]{FD6864} \textbf{0.6595}} \\ \hline\end{tabular}}
\end{table*}

\begin{figure*}[hbt!]
\centering
    \subfloat[SAR\label{1a}]{%
       \includegraphics[width=0.14\linewidth]{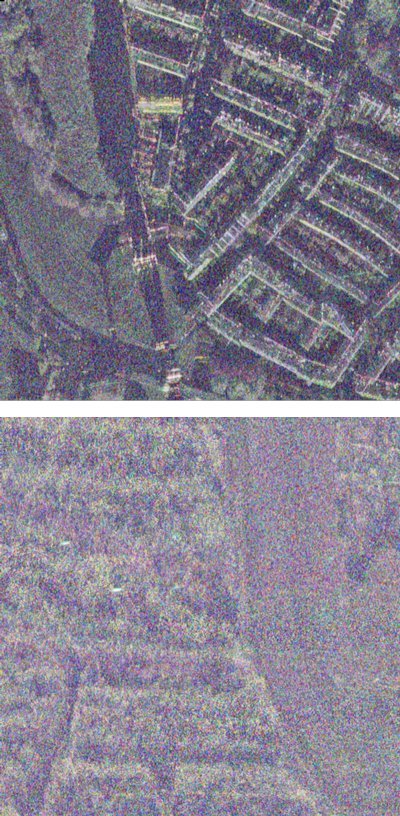}}
    \hfill
    \hspace{-15pt}
    \subfloat[pix2pix\label{1b}]{%
       \includegraphics[width=0.14\linewidth]{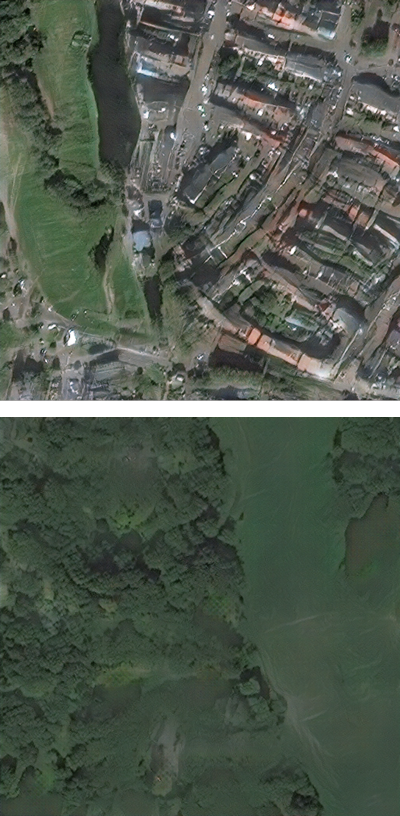}}
    \hfill
    \hspace{-15pt}
    \subfloat[cycleGAN\label{1c}]{%
       \includegraphics[width=0.14\linewidth]{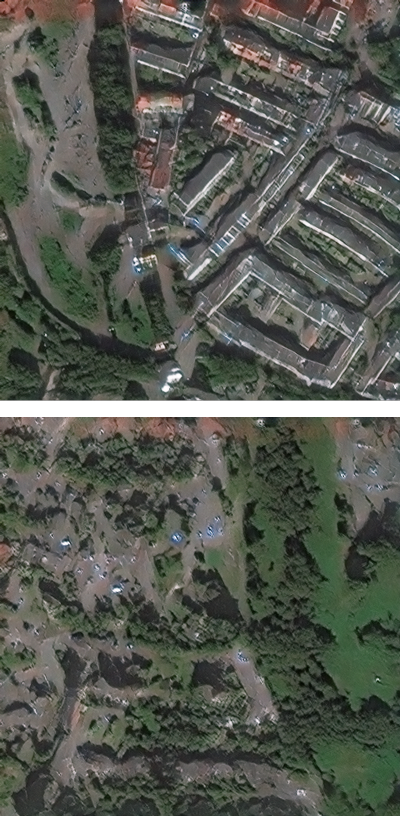}}
    \hfill
    \hspace{-15pt}
    \subfloat[cLDM\label{1d}]{%
       \includegraphics[width=0.14\linewidth]{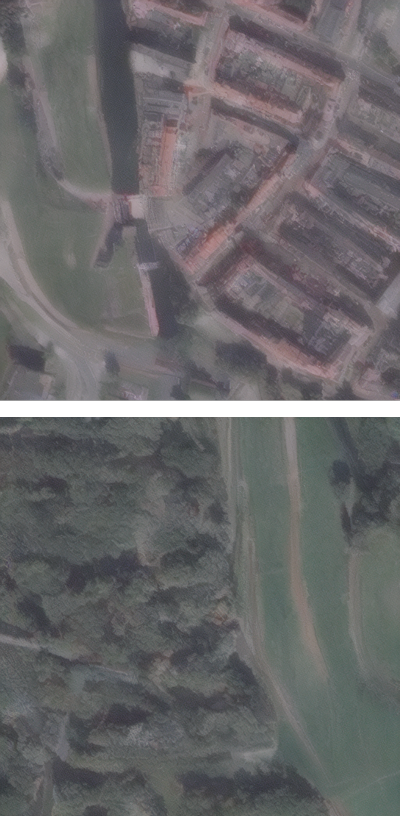}}
    \hfill
    \hspace{-15pt}
    \subfloat[BBDM\label{1e}]{%
       \includegraphics[width=0.14\linewidth]{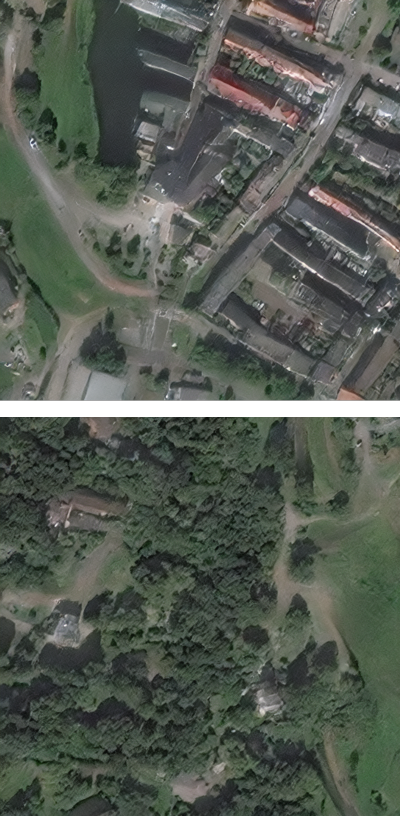}}
    \hfill
    \hspace{-15pt}
    \subfloat[cBBDM (Ours)\label{1f}]{%
       \includegraphics[width=0.14\linewidth]{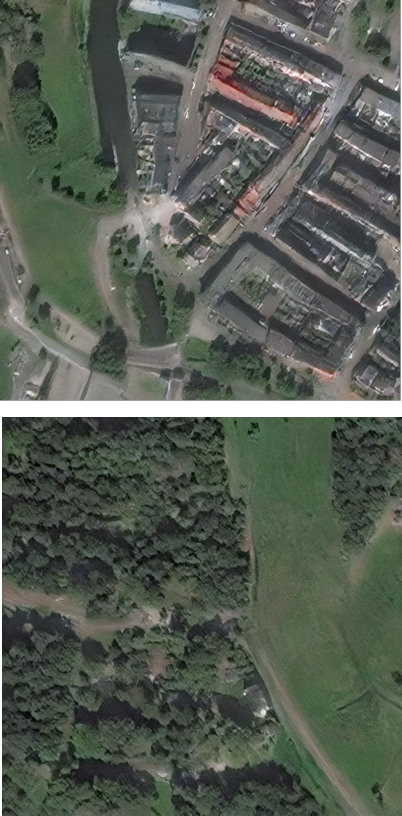}}
    \hfill
    \hspace{-15pt}
    \subfloat[optical\label{1g}]{%
       \includegraphics[width=0.14\linewidth]{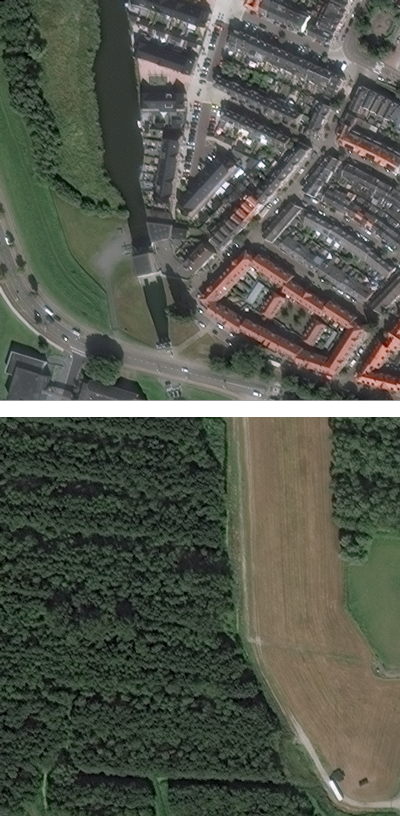}}
    \hfill
    \hspace{-15pt}
    \caption{Results of VHR SAR to optical image translation using different methods. The first row shows SAR to optical image translation for an urban scene. The second row shows translation for trees and bare land.}
    \label{fig:Qualitative comparison of translated images from different methods}
\end{figure*}

This approach enables cBBDM to produce optical-like images that refer to key features of the input SAR data more faithfully while preserving the directly mapping benefits of the original BBDM framework. The model can better preserve structural details and textures by leveraging SAR-specific features during the translation process. It potentially leads to more accurate translations and efficient training and inference.

\section{EXPERIMENTS AND RESULTS}
\subsection{Data Set}
We conducted experiments on the MSAW dataset \cite{MSAW_SpaceNet6} which consists of 0.5m VHR SAR and optical images captured by Capella Space X-band quad-pol SAR and WorldView-2 optical satellite, respectively. This dataset offers 3,401 geo-referrenced SAR and optical imagery pairs which are tiled to 900x900 pixels (450m x 450m) including zero-intensity background.

One or more 512x512 images were extracted from within each tile removing zero-intensity background. Since the paired sets overlap with each other, we split them into train and validation sets based on longitude to ensure that validation is conducted on unseen data. We created 3-channel false color composite with VV, VH, HH polarization and applied histogram equalization using CLAHE algorithm. We did not apply any noise reduction algorithm.

\subsection{Evaluation Metrics}
We used several metrics to assess image translation performance of our proposed method and to compare it with conditional LDM and GAN-based models. We utilized three different Learned Perceptual Image Patch Similarity (LPIPS) scores and Fréchet Inception Distance (FID), to evaluate perceptual similarity and distribution similarity between generated and authentic optical images, respectively. We further included Spectral Angle Mapper (SAM) and Color Histogram Distance (CHD) for spectral consistency evaluation. SAM measures the angular difference of spectral patterns and CHD measures the global difference of color histograms between two images. Additionally, to evaluate spatial quality, we utilized Complex Wavelet Structural Similarity (CW-SSIM) index, which measures structural coherence in the complex wavelet domain, and Chrominance Feature Similarity (FSIMc) index, which effectively captures key structural features such as edges and textures in the image\cite{huang2024generative}.

\subsection{Implementation Details}
We adopted a pre-trained Vector Quantized-Variational Auto Encoder(VQ-VAE) \cite{van2017neural}, a generative model that learns discrete latent representations by using a codebook to effectively reduce the pixel space dimensions by a factor of 4 in both the height and width directions. We trained the proposed model for 48 epochs using an NVIDIA A6000 GPU, with a minibatch size of 1 and Adam optimizer of 1e-4 learning rate. Data augmentation was implemented using horizontal flips during training. To ensure fair comparison, we applied same optimizer and augmentation techniques when training the conditional LDM and GAN-based models. These comparative models were trained using their default minibatch size and learning rates, which are widely adopted in the existing literature. The conditional LDM was fine-tuned from the pre-trained model by \cite{LDM}. For evaluation, we selected the checkpoint of each model that recorded the minimal loss on the validation dataset.

\subsection{Results and Analysis}
To evaluate our proposed method, we conducted comparisons with widely-used GAN-based models in the field of SAR to optical image translation, specifically Pix2Pix \cite{pix2pix} and CycleGAN \cite{CycleGAN}. In addition, we tested conditional LDM \cite{LDM} which is a representative Conditional Diffusion Model to assess the effectiveness of the BBDM framework based on Brownian bridge. Furthermore, we also tested original BBDM without any condition for ablation study.
\subsubsection{Quantitative Results}
Table \ref{table:Quantitative comparison of different methods} presents the quantitative comparison between ground truth optical images and translated optical-like images. Our proposed model significantly outperforms other models in all metrics. The consistently superior LPIPS scores demonstrate our model's robustly high perceptual quality. The best FID score suggests our model most accurately reconstructs the target distribution. Beyond perceptual metrics, our method also shows the best quality in spectral and structural consistency. The lowest SAM and CHD values indicate our model’s capability to maintain spectral characteristics and color distributions of authentic optical images. The highest CW-SSIM value confirms our approach's ability to maintain structural coherence. The best FSIMc score demonstrates that our method effectively maintains texture details and local structural features. Our model's superior performance in deep learning-based metrics like LPIPS and FID demonstrates its potential for deep learning-based downstream applications such as synthetic pretraining. Simultaneously, the best structural and spectral similarity metrics suggest that our approach can effectively assist SAR image interpretation by translating complicated SAR data into visually intuitive optical images.

\subsubsection{Qualitative Results}
Fig. \ref{fig:Qualitative comparison of translated images from different methods} illustrates translated optical-like images from various models. The first row shows SAR to optical translation results for a complex urban scene. Even though the generated images from Pix2pix initially appear satisfactory, closer inspection reveals significant loss of structural detail. CycleGAN failed to preserve surface boundaries and emphasized building side walls excessively. Although conditional LDM distinguished surfaces well, it produced a bit noisy images due to its excessive dependence on the given condition. In contrast, our conditional BBDM demonstrates superior reconstruction. It produced clearly discernible surfaces. Our approach outperforms the basic BBDM which missed some detailed information.

The second row shows SAR to optical translation results for a scene with trees and bare land. Our conditional BBDM successfully translated this challenging scene and clearly distinguished land and trees despite the potential confusion of low-intensity bare land with speckle noise. This demonstrates the robustness of proposed method against speckle noise. It generated distinguishable optical-like images from SAR images without any prior noise reduction. Although the colorization leans towards green tones, it outperforms other methods which produced indistinguishable surfaces with numerous artifacts. These results highlight the effectiveness of conditional BBDM in generating high quality and detailed optical-like images from SAR inputs across various scene types.

\section{Conclusion}
In this letter, we have proposed a new conditional BBDM framework for VHR SAR to optical image translation. Unlike the original BBDM which operates solely in latent space, our approach combines the benefits of Brownian Bridge mapping in latent space with explicit spatial conditioning from the pixel domain. With this integration, our proposed method achieved high quality and cost-effective translation for large VHR images. Quantitative and qualitative assessments indicate that proposed method generates impressive optical-like images and reveals its practical potential for SAR imagery interpretation. As an alternative to GAN-based approaches, our work contributes to improving SAR image interpretation and narrowing the gap between SAR and optical remote sensing modalities. This development potentially leads to improved applications. However, our study has limitations that warrant further investigation. The current research lacks comprehensive ablation studies exploring different spatial feature extraction methods and conditioning mechanisms. Therefore, in future work, we aim to supplement these studies and conduct architectural refinements for further improvement.

\bibliographystyle{IEEEtran}
\bibliography{references}
\end{document}